\documentclass[12pt]{article}

\usepackage{amsmath,amssymb,amsthm,graphicx,latexsym}

\newcommand{\ed}{\end{document}}
\newcommand{\be}{\begin{equation}}
\newcommand{\ee}{\end{equation}}
\newcommand{\pa}{\partial}

\begin{document}
\begin{center}
\large{\textbf{\textbf{Spontaneous Symmetry Breaking and Landau Phase Transition
in Horava Gravity}}}\\
\end{center}
\begin{center}
Sudipta Das\footnote{E-mail: sudipta.das\_r@isical.ac.in} and
Subir Ghosh\footnote{E-mail: sghosh@isical.ac.in}
\\
Physics and Applied Mathematics Unit, Indian Statistical
Institute\\
203 B. T. Road, Kolkata 700108, India \\
\end{center}\vspace{1.5cm}

\begin{center}
{\textbf{Abstract}}
\end{center}

\vskip .6cm
Presence of higher derivative terms in the Horava model of gravity can generate
an instability
in the Minkowski ground state. This in turn leads to a space dependent vacuum 
metric with a length scale determined by the higher derivative coupling
coefficient. The translation invariance is spontaneously  broken in the process. The phenomenon
is interpreted as a form of Landau liquid-solid phase translation.
The (metric) condensate  acts
as a source that modifies the Newtonian potential below the length scale but keeps it 
unchanged for sufficiently large distance.

\vskip 1.5cm

{\bf{Introduction and Formalism :}}
Horava Gravity (HG) \cite{hor} has created a lot of recent interest but
surprisingly an intriguing consequence of the all important higher derivative
extension terms in HG has gone unnoticed: possibility of an instability in the
conventional flat metric ground state leading to a 
richer structure of the vacuum. This is a form of the celebrated Landau
liquid-solid phase transition \cite{lan}
of the spacetime (only space in the present context) itself. In the present work
we show that this is indeed 
feasible and provide a preliminary study of some of its immediate effects.

An important lesson that we have learnt from Horava's \cite{hor} novel proposal
of a UV complete model of
Gravity is that Lorentz invariance is not that sacred in a theory. (See
\cite{vis} for a lucid exposition of the perspective regarding Lorentz
invariance violation.) It is apparently admissible even for a fundamental theory
of Gravity, such as the Horava model, to be Lorentz non-invariant at ultra high
energy or short distance, at $\sim$
trans-Planckian regime, so long as Lorentz invariance and Einstein General
Relativity (GR) is recovered at observable scales of low energy (or long
distance). This has brought about a paradigm shift in the way of thinking of
High Energy Physicists but is quite conventional for Condensed Matter Physicists
where maintenance of Lorentz invariance
is not an issue at all. Indeed, it was quite well known that higher derivative
covariant terms, (generically functions of $R$, the Ricci scalar),
in Einstein gravity improves the UV behavior but brings in insurmountable ghost
problems \cite{ste}. To overcome this in Horava Gravity (HG) only higher order
{\it{spatial}} derivative terms are kept.
These explicitly Lorentz breaking terms do not introduce ghosts and at the same
time can cure the UV divergence problems. However the jury is still out on the
question of whether
HG is the true theory of gravity since the loss of Lorentz and
diffeomorphism symmetries induces a spurious
degree of freedom beside the spin $2$ graviton \cite{pang}. Several ways to
improve the Horava model have also been suggested \cite{puj}.

On the other hand, the higher derivative nature of the  Horava model opens up a
completely new line of
thought that is of interest to us: instability in the flat metric ground state.
This can lead to a 
Spontaneous Symmetry Breaking (SSB) as regards to translation symmetry only.
{\it{This phenomenon is a liquid-solid type of phase transition
in the space(time) itself a la Landau \cite{lan}}}. A short distance length
scale is generated in the process
of transition from a homogeneous ($\sim$ liquid like) phase ground state to an
inhomogeneous ($\sim$ solid like) condensate phase ground state in
spacetime. Hence our work strengthens the claim of HG as a viable model
for Quantum Gravity because the physics below this scale is affected while GR is
recovered for distances sufficiently above this scale. 
The more ambitious project is to construct a true spacetime crystal
where translation as well as rotation symmetries are lost.
The obvious motivation is to build a suitable spacetime for Quantum Gravity
which is expected to have a short distance ($\sim$ Planck) scale {\footnote{
The attempts so far made are some what ad-hoc: (i)
Non-Commutative (NC) geometry framework \cite{ncgr} that
exploits Seiberg Witten map \cite{sw} and incorporates NC
corrections on GR. \\
(ii) A cutoff length scale in gravity was introduced \cite{lin} by mapping the
slice of
spacetime into the phase space of quantum mechanics of fermions. The length
scale and noncommutativity parameter in gravity map to the Planck's constant in
quantum mechanics.\\
(iii) A granular spacetime is considered in
\cite{Mod}.}}. Fortunately we know explicitly how to proceed in Condensed
Matter Physics \cite{alex}, where the physics is naturally non-relativistic.
Since we are in the Horava framework of gravity where Lorentz
invariance is explicitly broken, formalisms exploited in (non-relativistic)
Condensed Matter
Physics should be applicable in the present
scenario as well. We mention an earlier work by us 
\cite{dgh} where a length scale was introduced in the metric
in a higher derivative Lorentz covariant Gravity theory via SSB. However, the
new non-trivial ground state
broke the Lorentz invariance explicitly and furthermore the parent model has the
well known ghost problem. In 
this sense HG serves perfectly as the laboratory to test this new idea since HG
is free of ghosts and is not a Lorentz invariant theory to begin with (recall
that the higher
derivatives are only in the spatial sector).

In a series of pioneering works Alexander and Mctague \cite{alex} were able to
construct a solid crystalline lattice (where translation and rotation
symmetries are lost), from a liquid phase (where the symmetries
are intact), through SSB. This analysis explained the fact that $BCC$ lattice
structure is favored for metals
undergoing liquid-solid transition, essentially in a model independent way. In
order to achieve this the condensate has to have additional structure,
{\it{i.e.}}
its VEV can not be a constant that we generally encounter
in Particle Physics. Rather, the crucial feature of the condensate is that the
Fourier transform of non-zero VEV of condensate minimizing the free
energy must have support at a {\it{non-vanishing momentum}}. Evidently, as is
explained below, this requirement translates, in the coordinate space, to the
parent lagrangian undergoing SSB, as having
{\it{higher (at least fourth) order derivative}} terms that are quadratic in the
field. The field in question is identified with the difference between the
(inhomogeneous) density for solid and (constant) density of liquid, that acts as
the order parameter. The free energy is expanded around the
higher symmetry liquid phase and it should be mentioned that to form a proper
crystal lattice one needs terms, at least of third and fourth order terms in the
order parameter, in the expansion. However, the latter does not concern us in
the present work as we will restrict ourselves to {\it{higher derivative
quadratic terms}} in the HG action. The ideas developed in \cite{alex}
were applied in High Energy Physics
in the context of string compactification by \cite{rab}.

{\bf{SSB of Higher Derivative Scalar Theory:}} As a warm up exercise we
consider a toy model consisting of a
scalar field theory with higher
spatial derivative terms,
\be
S = \int  d^4 x \left[ \frac{1}{2} \phi (-\partial_0^2+\partial_i^2) \phi -
\frac{\alpha^2}{4} \phi (\partial_i^2)^2 \phi - V(\phi)\right]. \label{hdlag}
\ee
In fact one can think of (\ref{hdlag}) as obtained from
$$
S = \int  d^4 x \left[ \frac{1}{2} \phi (-\partial_0^2+\partial_i^2) \phi -
\frac{\alpha^2}{4} \phi (-\partial_0^2+\partial_i^2)^2 \phi - V(\phi)\right]$$ by
simply
leaving out the time derivatives in $\alpha ^2$ term. This is in the
same spirit as obtaining the Horava
model with higher order spatial Ricci scalar terms from a covariant higher
derivative gravity model containing higher orders of the full Ricci scalar.
Furthermore, as we will see
later, after some simplifying (but not unconventional)
restriction on the metric fluctuation, the Horava model will be structurally
identical to this toy model. In the present preliminary analysis we drop the
potential $V(\phi )$ as does not play any role since we are concerned only with
ground state solution obtained by minimizing the kinetic term. Minimizing
$V(\phi )$ restricts the solution, obtained by minimization of the kinetic
energy, still further.
Explicit structure of $V(\phi )$, such as with cubic and higher order
terms, becomes crucial in building the
lattice in momentum space of $\phi $ (see \cite{alex} for more details).

We are working in flat Minkowski spacetime with the metric $\eta_{\mu \nu}
\equiv
diag(-1, 1, 1, 1)$. The energy is given by
\be H = \int d^3x~[\frac{1}{2} \dot{\phi}^2 + \frac{1}{2} (\partial_i
\phi)^2 +
\frac{\alpha^2}{4} \phi (\partial_i^2)^2 \phi ]. \label{t00}
\ee
As explained above we have dropped $V(\phi )$ from further consideration.

Let us now minimize the energy. It is natural to consider the ground state to be
static. Using the Fourier decomposition $$ \phi(x) =
\frac{1}{(2\pi)^3}\int \phi(k) e^{i k.x} dk,$$  the ground state energy can be
written in the momentum space as
\be H =
-\frac{1}{2} k^2 \phi (\vec k) \phi (-\vec k)+ \frac{\alpha^2}{4}
(k^2)^2 \phi (\vec k) \phi (-\vec k). \label{t00static} \ee
Clearly $H[k^2 =0]=0$ but $H[k^2 =1/\alpha
^2]=-\frac{1}{4\alpha ^2}$ showing that the minimum of kinetic energy
(\ref{t00static}) occurs at $k^2 = \frac{1}{\alpha^2}$.
Here we used the notation $|\vec k|^2 = k^2$.

From  Figure 1, one can clearly see that the higher derivative
term shifts the position of the energy minimum from $k^2=0$ to
$k^2=1/\alpha ^2$. Subsequently translational invariance is broken in the
higher derivative theory \cite{alex,rab}.

\begin{figure}[htb]
{\centerline{\includegraphics[width=9cm, height=6cm] {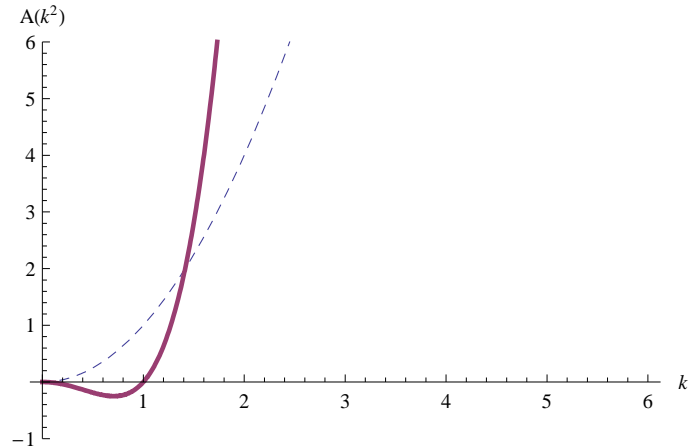}}}
\caption{{\it{Plot of $A(k^2)=k^2$
vs $k$ and $A(k^2)= - k^2(1 -\frac{\alpha ^2}{2}
k^2)$ vs $k$ where
the energies are respectively given by $H(k^2) = \phi(k) \phi(-k) A(k^2)$
and $H(k^2)=\phi(k) \phi(-k) A(k^2)$.
The specific form of the function $A(k^2)$ is determined
by the underlying microscopic theory \cite{rab}.
Translational symmetry is present in the first case (dashed line)
as the minima of $A(k^2)$ occurs at $k^2 = 0$. Translational symmetry is broken
in the second case (thick line)
as the minima of $A(k^2)$ occurs at $k^2 \neq 0$.}}} \label{fig1}
\end{figure}

The condensate value of $<\phi (x)>$, (that
minimizes the energy), has a nontrivial coordinate dependence
which can be determined as follows: \be <\phi(\mid x\mid)> =\frac{1}{(2\pi)^3}
\int  d^3k \delta(k^2-\frac{1}{\alpha ^2}) C(\alpha k) e^{i k.x} $$$$
=\frac{1}{(2\pi) ^2}\int d\theta~ sin\theta \int dk~k^2
\frac{1}{2k}\delta(k-\frac{1}{\alpha })C(\alpha k)e^{i k.x}
$$$$=\frac{C}{2\alpha (2\pi )^2}\int _0^{\pi }
d\theta ~sin\theta 
e^{(i\mid
x\mid cos\theta )/\alpha} $$$$ 
=\frac{C}{(2 \pi)^2} 
\frac{ sin(\mid x\mid /\alpha)}{|x|}, \label{phicon} \ee
where $C$ turns out to be a dimensionless number.
In the above expression (\ref{phicon}), notice that although the
condensate $\phi (\mid x\mid )$ has become explicitly
$x$-dependent, indicating that the translation 
invariance is lost,  the space rotational symmetry is preserved. This is
because in the Fourier integral the restriction was to integrate on the
spherical surface $k^2=1/\alpha ^2$
that does not affect rotational symmetry. For large $\alpha $ energy $H$ in
(\ref{t00static}) is minimum for $k^2=0$ and the condensate $<\phi (\mid x\mid
)>$ in (\ref{phicon}) becomes $\mid x\mid$-independent constant.

It is customary to consider fluctuations above the condensate by
shifting the field. In the present instance this exercise will means
$\phi(x)~\rightarrow~\phi(x)-\phi(\mid x\mid)$
leading to a translation {\it{non}}-invariant (but
rotation invariant) theory. There is an underlying periodic nature of the
translation symmetry broken phase. The coupling coefficient $\alpha $ in the
higher
derivative term in the action is responsible for the inhomogeneity
and is present in the resulting lagrangian. We have not provided an explicit
expression for that. Hence we have succeeded in introducing
a periodic structure in the deformed translation invariance through SSB and the
periodicity scale can be directly linked with the coupling coefficient of the
higher derivative term. We now repeat the same exercise on the
Horava model of Gravity in weak field approximation.

{\bf{SSB of Higher Derivative Horava Gravity:}}
As we have explained above, breaking of
translation invariance via SSB requires higher derivative terms in the action
and this motivates us to consider higher derivative gravity theories.
We take the higher derivative terms as perturbations on the linearized
Horava gravity action (which reproduces Einstein's GR theory in the low-energy
(IR) limit). As discussed before, Lorentz symmetry is already explicitly broken
in Horava theory.

Our long-term goal is to generate a spacetime lattice through SSB in the
way described in the previous section. However, in the present
article our target is more modest:
as a first step towards this objective, we intend to introduce a
short distance scale ($\sim$ Planck length) in the Horava action that
will break the translation invariance, without affecting the
rotational symmetry. Indeed, the extension of general relativity to NC
spacetimes \cite{ncgr} also aims to achieve that but, compared to this ad-hoc
procedure, our approach is much more basic and intuitive. We
introduce the deformation directly in the metric as a fluctuation and
the condensate plays the role of a (tensorial) order
parameter in the continuum-discrete spacetime phase transition. 
As we have extensively discussed before, this
requires the introduction of higher derivative terms in the action,
$R^{ij}R_{ij}$ and $(g^{ij}R_{ij})^2$.
Apart from the conventional $(g ^{ij}R_{ij})$-term, the above are
needed to ensure that the kinetic
term is minimized for a non-zero momentum.

We start with the  action $S$ 
\be
S=\int dt {\cal{L}}=\int dt d^3 x \sqrt{g} N (G^{ijkl}
K_{ij} K_{kl} + A R + B R_{ij} R^{ij} + C R^2 ) $$$$
= \int dt d^3 x \sqrt{g} N (K_{ij} K^{ij} - \lambda K^2
+ A R + B R_{ij} R^{ij} + C R^2 ).
\label{fullaction}
\ee

Here $g_{ij}$ is the spatial metric, $A, B, C$ are
three dimension-full parameters of the theory,
$N$ is the Lapse function, $R$ is the spatial Ricci
scalar and $K_{ij}$ is the extrinsic curvature defined as
\be
K_{ij}=\frac{1}{2 N}(\partial_0 g_{ij} - \nabla_i N_j - \nabla_j N_i).
\label{k}
\ee
with $N_i(x,t)$ is the Shift vector in ADM formalism \cite{adm} and
the generalized De Witt metric $G^{ijkl}$ is defined as
\be
G^{ijkl} = \frac{1}{2} (g^{ik}g^{jl} + g^{il}g^{jk}) - \lambda g^{ij} g^{kl}.
\label{dewitt}
\ee
For $\lambda = 1$, $A=1$ and $B=C=0$ Horava gravity reduces to
Einstein's gravity.

Throughout we will work in the weak field approximation with
\be
g_{ij}=\delta_{ij}+h_{ij}~~~,~~~N = 1+n~~~,~~~N_i=n_i. \label{per}
\ee

Under these perturbations (\ref{per}), the expressions for the
extrinsic curvature and the Ricci scalar turn out to be
\be
K_{ij}=\frac{1}{2}(\pa_0 h_{ij} - \pa_i n_j - \pa_j n_i)~~,~~
K=\delta^{ij}K_{ij}=\frac{1}{2}(\pa_0 h - 2 \pa_i n^i), $$$$
R_{ij}=\frac{1}{2}(\pa^k \pa_i h_{jk}+\pa^k \pa_j h_{ik}-\pa^2 h_{ij} -
\pa_i \pa_j h)~~,~~
R=\pa_i \pa_j h^{ij} - \pa^2 h. \label{kr}
\ee

Using the above expressions (\ref{kr}) and the relation
\be
\sqrt{g} R = \frac{1}{2} h_{ij}\left(-R^{ij}+\frac{1}{2}
\delta^{ij} R\right)
\label{gr}
\ee
in the action (\ref{fullaction}) we obtain the
lagrangian density ${\cal{L}}$ of second order in $h$:
\be
{\cal{L}}=\frac{1}{4}[\pa_0 h_{ij} \pa_0 h^{ij} - \lambda(\pa_0 h)^2
-4(\pa_0 n_i)(\pa_j h^{ij} - \lambda \pa^i h)
+ (4 \lambda -2)n_i (\pa^i \pa^j n_j) - 2 n_i \pa^2 n^i] $$$$
+ \frac{A}{4} h_{ij} (\pa^2 h^{ij} - 2 \pa_k \pa^i h^{jk}
+ 2 \pa^i \pa^j h - \delta^{ij} \pa^2 h)
+ A n (\pa_i \pa_j h^{ij} - \pa^2 h)
+ C (\pa_i \pa_j h^{ij} - \pa^2 h)(\pa_k \pa_l h^{kl} - \pa^2 h)
$$$$ + \frac{B}{4} (\pa^k \pa_i h_{jk} - \pa^2 h_{ij}
+ \pa^k \pa_j h_{ik} - \pa_i \pa_j h)
(\pa_l \pa^i h^{jl} - \pa^2 h^{ij}
+ \pa_l \pa^j h^{il} - \pa^i \pa^j h). \label{lag}
\ee

In the above expression (\ref{lag}) and throughout the rest of our paper, we
used the following notation:
$$ h = \delta^{ij} h_{ij}~~~~,~~~~\pa^2 = \pa_i \pa^i = \delta^{ij}
\pa_i \pa_j.$$
From this lagrangian (\ref{lag}) we obtain the conjugate momenta as
\be
p \equiv \frac{\pa {\cal{L}}}{\pa (\pa_0 n)} = 0~~~,~~~
p^i \equiv \frac{\pa {\cal{L}}}{\pa (\pa_0 n_i)} =
-(\pa_j h^{ij} - \lambda \pa^i h), \label{p}
\ee
\be
\pi^{ij} \equiv \frac{\pa {\cal{L}}}{\pa (\pa_0 h_{ij})} =
\frac{1}{2} (\pa_0 h^{ij} - \delta^{ij} \lambda (\pa_0 h)).
\label{pi}
\ee

Taking the trace of the relation (\ref{pi}), we can write
$\pa_0 h_{ij}$ in terms of $\pi_{ij}$ as
\be
\pa_0 h^{ij} = 2 \left(\pi^{ij}+\frac{\lambda}{1 - 3 \lambda}
\delta^{ij} \pi \right). \label{delh}
\ee

Using the relations (\ref{p}), (\ref{pi}), (\ref{delh}) in (\ref{lag})
we get the Hamiltonian density,
\be
{\cal{H}} = p^i (\pa_0 n_i) + \pi^{ij} (\pa_0 h_{ij}) - {\cal{L}} $$$$
= \pi_{ij} \pi^{ij} - \frac{\lambda}{3 \lambda - 1} \pi^2 -
\frac{1}{2} (2 \lambda - 1) n_i (\pa^i \pa^j n_j) + \frac{1}{2}
n_i \pa^2 n^i $$$$ - \frac{1}{4} h_{ij} (\pa^2 h^{ij} - 2 \pa_k \pa^i h^{jk}
+ 2 \pa^i \pa^j h - \delta^{ij} \pa^2 h)
- A n (\pa_i \pa_j h^{ij} - \pa^2 h)
- C (\pa_i \pa_j h^{ij} - \pa^2 h)(\pa_k \pa_l h^{kl} - \pa^2 h)
$$$$ - \frac{B}{4} (\pa^k \pa_i h_{jk} - \pa^2 h_{ij}
+ \pa^k \pa_j h_{ik} - \pa_i \pa_j h)
(\pa_l \pa^i h^{jl} - \pa^2 h^{ij}
+ \pa_l \pa^j h^{il} - \pa^i \pa^j h). \label{ham1}
\ee
Using the momentum constraints and the gauge $n_i=0$,
the Hamiltonian (\ref{ham1}) now reduces to
\be
{\cal{H}} = \pi_{ij} \pi^{ij} - \frac{1}{4} h_{ij} (\pa^2 h^{ij} - 2 \pa_k \pa^i
h^{jk} + 2 \pa^i \pa^j h - \delta^{ij} \pa^2 h)
$$$$ - \frac{B}{4} (\pa^k \pa_i h_{jk} - \pa^2 h_{ij}
+ \pa^k \pa_j h_{ik} - \pa_i \pa_j h)
(\pa_l \pa^i h^{jl} - \pa^2 h^{ij}
+ \pa_l \pa^j h^{il} - \pa^i \pa^j h). \label{hamiltonian}
\ee

In the static limit $\pi_{ij}=0$ and with $h_{ij} = \delta_{ij} h$ for the sake of convenience, the
Hamiltonian
(\ref{hamiltonian}) can be written in the Fourier space as
${\cal{H}}_{static}(p)$,
\be
{\cal{H}}_{static}(p) = -\frac{p^2}{6} h(p) h(-p) \left(1-\frac{17 B}{5} p^2
\right).
\label{hstatic} \ee
The minimum of this (\ref{hstatic}) occurs at
$$ p^2 = \frac{1}{\alpha^2} = \frac{5}{34 B}.$$
In a similar way as in (\ref{phicon}) we have the explicit form of $f(r)$ as
\be f(r) =  \frac{sin(r / \alpha)}{r}, \label{f} \ee
where $r=|x|$ stands for the radial coordinate.

Now, to consider the SSB effects, let us consider fluctuations
${\tilde{h}}_{\mu \nu}$ above the condensate,
\be  h_{ij} = \delta_{ij} f(\mid {\vec x} \mid) + {\tilde{h}}_{ij}(x)
$$$$ h_{00} = \tilde h_{00}= - 2 \phi~~~~,~~~~h_{0i}=0, \label{metric} \ee
where $f(\mid {\vec x} \mid)$ is a purely spatial function.
The $f(\mid {\vec x} \mid)$ term, which is exactly of the same form
as in (\ref{phicon}), comes from the energy
minimization arguments (previously described in the scalar theory case)
since we do not want the presence of a constant vector
in the theory as it would break the rotational invariance, and
${\tilde{h}}_{\mu \nu}$ is treated as the order parameter.
{\footnote{It is now clear that to make a full
fledged spacetime ``crystal'' one requires third and fourth terms
in $h_{\mu\nu }$, (as explained in \cite{alex,rab} for
liquid-solid transition), in the action which will considerably
complicate the model. We are not considering this additional terms
in the present work.}}

In terms of the metric (\ref{metric}), the lagrangian (\ref{lag}) becomes
\be
{\cal{L}}=\frac{1}{4}[\pa_0 {\tilde{h}}_{ij} \pa_0 {\tilde{h}}^{ij} -
\lambda(\pa_0 {\tilde{h}})^2]
+ \frac{A}{4} {\tilde{h}}_{ij} (\pa^2 {\tilde{h}}^{ij} - 2 \pa_k \pa^i
{\tilde{h}}^{jk}
+ 2 \pa^i \pa^j {\tilde{h}} - \delta^{ij} \pa^2 {\tilde{h}}) $$$$
+ A n (\pa_i \pa_j {\tilde{h}}^{ij} - \pa^2 {\tilde{h}}) + C (\pa_i \pa_j h^{ij}
- \pa^2 h)(\pa_k \pa_l h^{kl} - \pa^2 h)
$$$$ + \frac{B}{4} (\pa^k \pa_i {\tilde{h}}_{jk} - \pa^2 {\tilde{h}}_{ij}
+ \pa^k \pa_j {\tilde{h}}_{ik} - \pa_i \pa_j {\tilde{h}})
(\pa_l \pa^i {\tilde{h}}^{jl} - \pa^2 {\tilde{h}}^{ij}
+ \pa_l \pa^j {\tilde{h}}^{il} - \pa^i \pa^j {\tilde{h}}) $$$$
+\frac{A}{2} {\tilde{h}}^{ij} (\pa_i \pa_j f - \delta_{ij} \pa^2 f) - 4 C
{\tilde{h}}^{ij}
(\pa_i \pa_j \pa^2 f - \delta_{ij} (\pa^2)^2 f) - \frac{3 B}{2} {\tilde{h}}^{ij}
(\pa_i \pa_j \pa^2 f - \delta_{ij} (\pa^2)^2 f). \label{lag2}
\ee
Varying this lagrangian (\ref{lag2}) with respect to the fields $n$ and
${\tilde{h}}^{ij}$ we get the following equations of motion
\be
\pa_i \pa_j {\tilde{h}}^{ij} - \pa^2 {\tilde{h}} - 2 \pa^2 f = 0, \label{eomn}
\ee

\be
\frac{1}{2}(\pa_0 \pa_0 {\tilde{h}}_{ij} - \lambda \pa_0 \pa_0 {\tilde{h}})
+ \frac{A}{2} (\pa^2 {\tilde{h}}_{ij} - 2 \pa^k \pa_i {\tilde{h}}_{jk}
+ \pa_i \pa_j {\tilde{h}} +  \delta_{ij} \pa^k \pa^l {\tilde{h}}_{kl} -
\delta_{ij} \pa^2 {\tilde{h}}) $$$$
+ A (\pa_i \pa_j n - \delta_{ij} \pa^2 n) +
2 C (\pa_i \pa_j \pa_k \pa_l {\tilde{h}}^{kl} - \pa^2 \pa_i \pa_j {\tilde{h}}
- \delta_{ij} \pa^2 \pa_k \pa_l {\tilde{h}}^{kl} + \delta_{ij} (\pa^2)^2
{\tilde{h}}) $$$$
+ \frac{B}{2} (2 \pa_i \pa_j \pa_k \pa_l {\tilde{h}}^{kl} - \pa^2 \pa^k \pa_i
{\tilde{h}}_{jk} - \pa^2 \pa^k \pa_j {\tilde{h}}_{ik} + (\pa^2)^2
{\tilde{h}}_{ij} - \delta_{ij} \pa^2 \pa_k \pa_l {\tilde{h}}^{kl} + \delta_{ij}
(\pa^2)^2 {\tilde{h}}) $$$$
+ \frac{A}{2} (\pa_i \pa_j f - \delta_{ij} \pa^2 f) - 4 C (\pa_i \pa_j \pa^2 f -
\delta_{ij} (\pa^2)^2 f) - \frac{3 B}{2} (\pa_i \pa_j \pa^2 f - \delta_{ij}
(\pa^2)^2 f) = 0. \label{eomh} \ee
Note that the $f$-terms in (\ref{eomh}) and in (\ref{eomn}) are subsequently
acting
 as the components of a conserved source.

Following standard literature \cite{car}, we decompose
$\tilde{h}_{\mu \nu}$ as:
\be \tilde{h}_{00} = -2n = -2 \phi~~~,~~~\tilde{h}_{0i} = 0~~~,~~~\tilde{h}_{ij}
= 2s_{ij} - 2 \psi \delta_{ij}, \label{dh} \ee
where $s_{ij}$ is traceless.
Using the above decomposition (\ref{dh}) and considering the static limit (hence
dropping all time derivatives), the equations of motion (\ref{eomn}, \ref{eomh})
turn out to be:
\be
\pa^2 \psi = \frac{1}{2} \pa^2 f, \label{eomgreen}
\ee

\be
A (\pa_i \pa_j - \delta_{ij} \pa^2) (\phi - \psi) + A \pa^2 s_{ij}
+ \frac{A}{2} (\pa_i \pa_j - \delta_{ij} \pa^2) f + 8 C \pa^2 (\pa_i \pa_j -
\delta_{ij} \pa^2) \psi $$$$ - 4 C \pa^2 (\pa_i \pa_j - \delta_{ij} \pa^2) f
+ B (\pa^2)^2 s_{ij} - 3 B \delta_{ij} (\pa^2)^2 \psi - \frac{3 B}{2}
\pa^2 (\pa_i \pa_j - \delta_{ij} \pa^2) f = 0. \label{eomphs}
\ee
The
solution for $\psi$ is obtained from (\ref{eomgreen}) as
\be
\psi(r) = -\frac{GM}{r} + \frac{1}{2}\frac{sin(\frac{r}{\alpha})}{r},
\label{psi} \ee
where the first term in the right hand side of (\ref{psi}) is the usual term
as obtained in linearized Einstein Gravity and the second term
is the modification due to the condensate.
We now take the trace of both sides of the equation (\ref{eomphs})
to obtain the following relation between $\phi$ and $\psi$:
\be
\pa^2 [\phi - \psi + \frac{f}{2} + \frac{3 B}{4 A} \pa^2 f] = 0 $$$$
\Longrightarrow \phi = \psi - \frac{f}{2} - \frac{3 B}{4 A} \pa^2 f
\label{phipsi}
\ee
Using the explicit expression of $f(r)$ as given in (\ref{f}),
the expression for $\phi$ turns out to be
\be
\phi(r) =  -\frac{GM}{r} + \frac{3B}{4A\alpha
^2}\frac{sin(\frac{r}{\alpha})}{r}.
 \label{phi}
\ee
The modified Newtonian potential derived above in (\ref{phi}) is shown in Figure
2. One can 
see that the fluctuations die out for sufficiently large distance $r>> \alpha $
but can have 
non-trivial effects for $r\sim \alpha $.

\begin{figure}[htb]
{\centerline{\includegraphics[width=9cm, height=6cm] {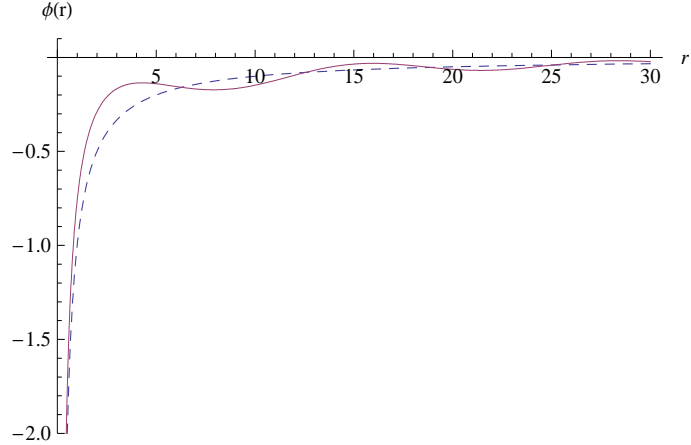}}}
\caption{{\it{Plot of the usual Newtonian potential (thin line in the plot)
and the modified Newtonian potential as described in previous paragraph (dashed line in the plot)
vs $r$ with $\alpha=2$.}}} \label{fig2}
\end{figure} 
\vskip .6cm

{\bf{Conclusion:}}~~~
To conclude, we have shown that higher
derivative terms in the  Horava model of gravity can lead to an instability and
a spontaneous breaking of
translation symmetry of the space takes place. Hence ground state gets modified
from the conventional
Minkowski vacuum metric to a space dependent form. This is an 
example of  Landau type of liquid-solid phase transition occurring in the space
itself. A length scale also appears in the resulting metric in a natural way.
The Newtonian potential
is recovered for distances large compared to this scale. Consequences of this
novel form of spatial metric
is worth pursuing.

{\it{Acknowledgments}}: It is a pleasure to thank Professor Chetan Nayek and
Professor Eliezer Rabinovici for suggestions and Professor Robertus Potting and
Professor Partha Mitra for correspondence and discussions. We also thank
Professor Pierro Nicolini and
Professor Hai Lin for informing us of their works in related areas.


\begin{thebibliography}{99}

\bibitem{hor} P. Horava, Phys. Rev. D 79 (2009) 084008
[arXiv:0901.3775].

\bibitem{lan} L. D. Landau, Phys. Z. Soviet II 26, 545 (1937).

\bibitem{vis} M. Visser, [arXiv:1103.5587].

\bibitem{ste} K. S. Stelle, Gen. Rel. Grav. 9 353 (1978).

\bibitem{pang} D. Blas, O. Pujolas, S. Sibiryakov,
JHEP 10 (2009) 029 [arXiv:0906.3046]; M. Li and Y. Pang, JHEP 08 (2009) 015
[arXiv:0905.2751]; M. Henneaux, A. Kleinschmidt and G. L.
Gomez, Phys. Rev. D 81 (2010) 064002 [arXiv:0912:0399].

\bibitem{puj} D. Blas, O. Pujolas, S. Sibiryakov,
Phys. Rev. Lett. 104 (2010) 181302 [arXiv:0909.3525]; J. Bellorin and A.
Restuccia,
[arXiv:1004.0055]; {\it{ibid}} J. Bellorin and A. Restuccia,
Phys. Rev. D 83 (2011) 044003 [arXiv:1010.5531]; S. Das and S. Ghosh,
[arXiv:1104.1975];
W. Donnelly and T. Jacobson, [arXiv:1106.2131].


\bibitem{ncgr} P. Nicolini, A. Smailagic and E. Spallucci, Phys. Lett. B 632,
547 (2006) [arXiv:gr-qc/0510112]. For a review see, P. Nicolini, Int. J. Mod.
Phys. A 24
(2009) 1229-1308 [arXiv:0807.1939].

\bibitem{sw} N. Seiberg and E. Witten, JHEP 9909 032 (1999)
[arXiv:hep-th/9908142].

\bibitem{lin} H. Lin, O. Lunin, J. Maldacena; JHEP 0410 (2004) 025
[arXiv:hep-th/0409174]; G. T. Horowitz, J. Polchinski, in {\it{Towards quantum
gravity}}, ed. D. Oriti, Cambridge University Press, [arXiv:gr-qc/0602037].

\bibitem{Mod} L. Modesto and P. Nicolini, Phys. Rev. D 81, 104040 (2010)
 [arXiv:0912.0220].

\bibitem{alex} S. Alexander and J. Mctague, Phys. Rev. Lett. 41 703 (1978); S.
Alexander, Symmetries and Broken Symmetries in Condensed Matter
Physics, ed. N. Boccara (IDSET Paris,1981) p. 141, and references therein;
Journal de Physique, Colloque C3, March 1985.

\bibitem{dgh} S. Das and S. Ghosh, [arXiv:1006.5774].

\bibitem{rab} E. Rabinovici, Spontaneous Breaking of Space-Time
Symmetries, {\it{Lect. Notes Phys. 737:573-605, 2008}}.
(arXiv:0708.1952); S. Elitzur, A. Forge and E. Rabinovici, Nucl. Phys. B 359
(1991) 581.


\bibitem{adm} R. L. Arnowitt, S. Desser and C. W. Misner, Gravitation: an
Introduction to Current Research (Wiley 1962), chapter 7, pp 227-265.

\bibitem{car} S. Carroll, {\it{Spacetime and Geometry, An Introduction to
General Relativity}}, Addison Wesley, 2004.

\end{thebibliography}
\end{document}